\def\cm{cm$^{-1}$}

\def\BaFeCoxAs{Ba\-(Fe$_{1-x}$\-Co$_x$)$_2$As$_{2}$}
\def\bfca{Ba\-(Fe$_{0.9}$\-Co$_{0.1})_2$As$_{2}$}

\def\tc{$T_{c}$}

\documentclass[prb,twocolumn,showpacs,superscriptaddress,amsmath,amssymb]{revtex4}
\usepackage[dvips]{graphicx}
\usepackage{graphicx}

\begin{document}

\title{Direct observation of a nodeless superconducting energy gap\\
in the optical conductivity of iron-pnictides}

\author{B. Gorshunov}
\affiliation{1.~Physikalisches Institut, Universi\"at Stuttgart, Pfaffenwaldring 57, 70550 Stuttgart, Germany}
\affiliation{Prokhorov Institute of General Physics, Russian Academy of Sciences,  Vavilov str. 38, 119991 Moscow, Russia}
\author{D. Wu}\email{dan.wu@pi1.physik.uni-stuttgart.de}
\affiliation{1.~Physikalisches Institut, Universi\"at Stuttgart, Pfaffenwaldring 57, 70550 Stuttgart, Germany}

\author{A. A. Voronkov}
\affiliation{Prokhorov Institute of General Physics, Russian Academy of Sciences,  Vavilov str. 38, 119991 Moscow, Russia}
\author{P. Kallina}
\affiliation{1.~Physikalisches Institut, Universi\"at Stuttgart, Pfaffenwaldring 57, 70550 Stuttgart, Germany}
\author{K. Iida}
\author{S. Haindl}
\author{F. Kurth}
\author{L. Schultz}
\author{B. Holzapfel}
\affiliation{IFW Dresden, Institute for Metallic Materials, P. O. Box 270116, 01171 Dresden, Germany}
\author{M. Dressel}
\affiliation{1.~Physikalisches Institut, Universi\"at Stuttgart, Pfaffenwaldring 57, 70550 Stuttgart, Germany}

\date{\today}

\begin{abstract}
The temperature-dependent optical reflectivity and complex transmissivity
of an epitaxially grown Ba(Fe$_{0.9}$Co$_{0.1}$)$_2$As$_2$ thin film were measured
and the optical conductivity and permittivity evaluated over a wide frequency range.
The opening of the superconducting gap $2\Delta_0 = 3.7$~meV
below $T_c\approx 20$~K is {\em directly} observed by a completely vanishing optical conductivity. The temperature and frequency dependent
electrodynamic properties of Ba(Fe$_{0.9}$Co$_{0.1}$)$_2$As$_2$ in the superconducting state agree well with the BCS predictions
with no nodes in the order parameter.  The spectral weight of the condensate $1.94\times 10^7~{\rm cm}^{-2}$ corresponds to a London penetration depth $\lambda_L=3600$~\AA.
\end{abstract}

\maketitle

\section{Introduction}
Soon after the discovery of superconductivity in iron-pnictides,\cite{Kamihara08} the epitaxial growth of LaFeAsO films was reported \cite{Hiramatsu08a,Backen08} and superconductivity observed in thin films of Co doped SrFe$_2$As$_2$.\cite{Hiramatsu08b,Maiorov09} By now the homogeneity of the films and the upper critical field have increased to make iron-pnictides interesting for technological applications.\cite{Haindl09,Kidszun09,Katase09}
In particular cobalt-doped BaFe$_2$As$_2$ seems to be suitable for producing
high-quality thin films which are stable in air,\cite{Katase09} can be template engineered,\cite{Lee09} or tuned in $T_c$ by epitaxial growths of strained films.\cite{Iida09}

Besides potential applications, thin films are advantageous for investigations of fundamental problems due to their large area, in particular if single crystals of sufficient quality, homogeneity, and size are limited. As far as optical experiments are concerned, only thin films give the opportunity to perform transmission measurements and in this way be much more sensitive to probe the electrodynamic properties of the normal and superconducting states.\cite{DresselGruner02,Dressel08}
Issues like the spectral weight distribution, the universal conductivity background, and in particular on the superconducting gaps, on states in the gap, nodes in the order parameter and quasi-particle relaxation are addressed by our optical experiments on \bfca\ thin films.

\section{Experimental Details and Results}

\begin{figure}
    \centering
        \includegraphics[width=0.7\columnwidth]{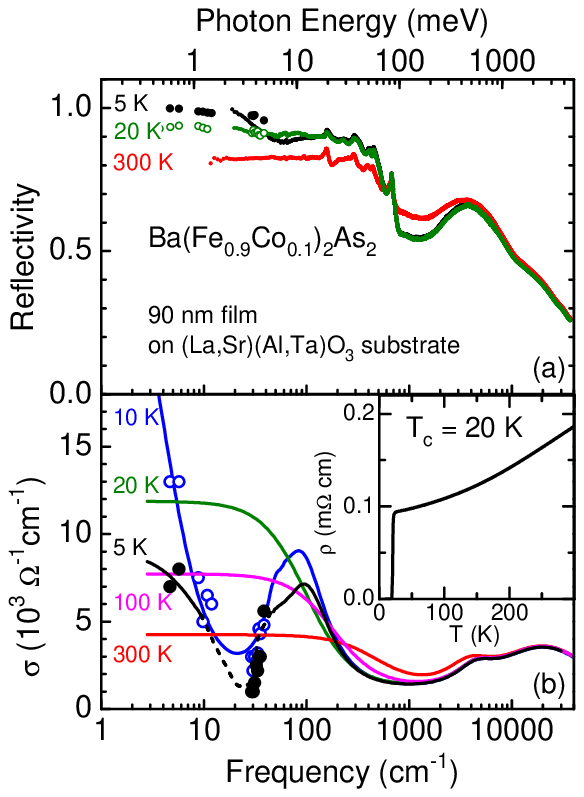}
    \caption{\label{fig:refcond1} (Color online) (a) Reflectivity of a 90~nm \bfca\ film on a 1~mm (La,Sr)(Al,Ta)O$_3$ substrate measured in a wide frequency range at various temperatures. The dots between 4 and 40~\cm\ are calculated from the transmission measurements.
(b)~Optical conductivity of \bfca\ obtained from the Drude-Lorentz analysis of the reflection and transmission spectra. The dots are directly calculated from measuring the transmission and phase by a Mach-Zehnder interferometer. The dashed part of the 5~K curve between 10 and 50~\cm\ indicates that a simple Lorentz shape does not mimic the superconducting gap properly since $\sigma(\omega)$ basically vanishes abruptly at 30~\cm. The inset shows the dc resistivity $\rho(T)$ of our sample with $T_c=20$~K.}
\end{figure}

\BaFeCoxAs\ films were deposited on a (001)-orientated
(La,Sr)(Al,Ta)O$_3$ substrate by pulsed laser deposition, where
the \bfca\ target was ablated with 248 nm KrF radiation under UHV
conditions.\cite{Iida09} The films grow with a very smooth surface
with an rms roughness better than 12~nm, as measured by atomic
force microscopy (AFM). The film thickness was monitored {\it in
situ} by a quartz balance, and finally measured by AFM and
ellipsometry to be $d=90$~nm. The phase purity was checked by
X-ray and EDS. Standard four-probe method was utilized to measure
the dc resistivity and determine the superconducting transition:
from the onset at 22~K with a transition width of 2~K, we have
chosen $T_c = 20$~K (inset of Fig.~\ref{fig:refcond1}).

Using different optical methods, we performed experiments in the frequency range from 4 to 35\,000~\cm\ and at various temperatures down to 5~K. In the THz range (4 to 40~\cm) the complex transmissivity (transmission coefficient amplitude and phase) was measured utilizing a Mach-Zehnder arrangement.\cite{Gorshunov05}
Between 20 and 15\,000~\cm\ the reflectivity was investigated by Fourier transform infrared spectroscopy; a gold mirror served as reference.  The spectra were extended up to the ultraviolet by  room-temperature ellipsometric data (6000 - 35\,000~\cm). In order to determine the properties of the film, we measured the optical parameters of a bare (La,Sr)(Al,Ta)O$_3$ substrate  over the entire frequency and temperature ranges.

In Fig.~\ref{fig:refcond1}(a) the optical reflectivity is plotted in a wide frequency range for selected temperatures. In particular in the far-infrared range the phonons of the substrate become obvious. For the further analysis we therefore employ a two-layer model that consists of the (La,Sr)(Al,Ta)O$_3$ substrate with thickness of 1.023~mm and  optical parameters determined beforehand, covered by the thin film of \bfca. Using Fresnel's equations\cite{DresselGruner02} we then can analyze the intrinsic optical properties of the film. In the THz range (4 to 40~\cm) where data for the transmission and phase shift are available, a corresponding analysis was performed that allowed us to directly determine the values of dielectric permittivity and conductivity, with the experimental uncertainties strongly dependent on the values of $\epsilon(\omega,T)$ and $\sigma(\omega,T)$.\cite{Gorshunov05} The same model was used to evaluate the optical response of the film at higher frequencies.
Eventually a self-consistent fit of the conductivity and permittivity spectra at THz frequencies and the reflectivity spectra at higher frequencies yields the overall behavior as presented in Fig.~\ref{fig:refcond1}(b).

The properties of the \bfca\ film in the normal state are described by two Drude terms, a narrow $\sigma_N$ and a broad one $\sigma_B$, corresponding to two types of charge carriers as suggested in Ref.~\onlinecite{Wu10}. In addition, two Lorentz terms account for the interband transitions.
In the superconducting state, two additional Drude terms are introduced, one with a tiny scattering rate to model the $\delta$-function (Cooper pair response) obvious in the permittivity spectrum, and another term to describe the quasi-particle contribution to the below-gap conductivity. The optical response at energies around the superconducting gap are mimicked with two Lorentzians since an appropriate expression is missing.
The resulting curves $\sigma(\omega,T)$ are plotted in Fig.~\ref{fig:refcond1}(b) as a function of frequency for selected temperatures.

\begin{figure}
    \centering
        \includegraphics[width=0.7\columnwidth]{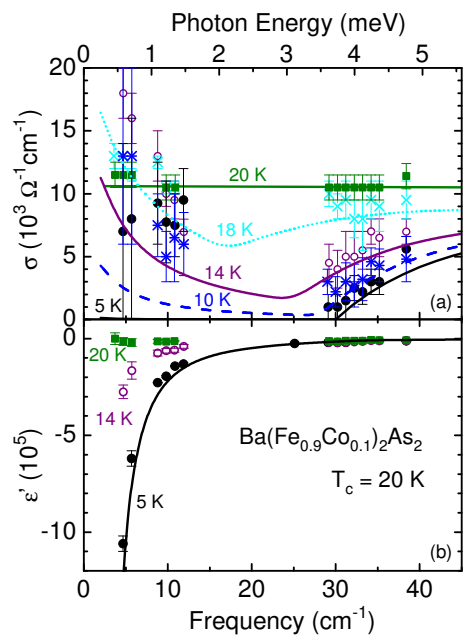}
    \caption{\label{fig:condeps1} (Color online) Optical properties of \bfca\ in the THz range where transmission and phase measurements directly yield the (a) conductivity and (b) permittivity spectra. The lines are calculated from the BCS theory assuming a complete opening of the gap at 30~\cm\ over the entire Fermi surface.}
\end{figure}

The most important finding of our investigation is the distinct
opening of the superconducting gap which is directly seen in the
drop of $\sigma(\omega,T)$ around 30~\cm\ upon cooling below \tc,
as depicted in Fig.~\ref{fig:condeps1}(a); for $T=5$~K the
conductivity  has completely vanished. The depletion extends up to
approximately 200~\cm. Due to remaining quasi-particles, the
conductivity becomes large below 10~\cm\ and even exceeds the
normal state value. The enormous drop of the low-frequency
permittivity $\epsilon^{\prime}(\omega,T)$ plotted in
Fig.~\ref{fig:condeps1}(b) evidences the inductive response of the
superconducting condensate.  It should be noted that the overall
conductivity of the \bfca\ film is identical to findings on single
crystals\cite{Wu10,Barisic10,Wu10b} and thus resembles the
intrinsic and general optical behavior of 122 iron-pnictides.

\section{Analysis and Discussion}
In the metallic state ($T>20$~K) the optical conductivity of \bfca\
has three major components: (i)~The infrared peaks at 4\,400 and  20\,800~\cm\ indicate interband transitions. (ii)~The background $\sigma_B \approx 1000~(\Omega {\rm cm})^{-1}$, which is best seen in the conductivity minimum around 1000~\cm, is  more or less temperature independent; we model it with a Drude term, albeit its roll-off cannot really be determined. (iii)~Most important is the narrow Drude component $\sigma_N(\omega)$, which grows and becomes sharper as $T$ decreases. Its spectral weight decreases by 15\%\
when the temperature is reduced from 300 to 20~K.
This corresponds to electronic correlation effects previously observed.\cite{Qazilbash09}

For $T<T_c$ the THz conductivity dramatically decreases below 100~\cm\ due to the opening of the superconducting gap at 30~\cm, as seen in Fig.~\ref{fig:refcond1}(b). The directly measured conductivity is displayed in more detail in Fig.~\ref{fig:condeps1}(a) together with calculations of the BCS-based Mattis-Bardeen model.\cite{Mattis58,Tinkham96,DresselGruner02} The best description is obtained for $2\Delta_0=3.7$~meV, corresponding to $2\Delta_0/k_BT_c\approx 2.1$; this value is considerably lower than expected from mean-field theory, but similar to what has been determined from reflection measurements of single crystals.\cite{Wu10,Wu10b} The shape of $\sigma(\omega)$ perfectly agrees with the BCS prediction for a simple $s$-wave superconductor with no indications of states or nodes in the gap. Previous optical investigations\cite{Li08,Hu09c} on hole-doped 122 iron-pnictides draw similar conclusions.

\begin{figure}
    \centering
        \includegraphics[width=0.7\columnwidth]{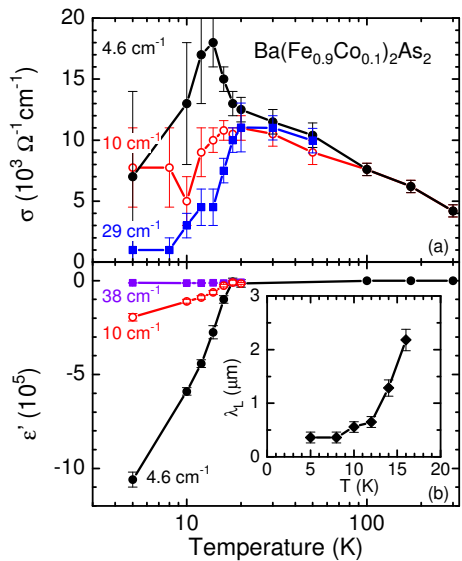}
    \caption{\label{fig:lambda1} (Color online) Temperature dependence of the low-frequency (a) optical conductivity $\sigma(\omega,T)$, (b) permittivity $\epsilon(\omega,T)$ and London penetration depth ($\lambda_L(T)$. For the lowest frequencies $\sigma(T)$ clearly exhibits a maximum below \tc.}
\end{figure}
Below 10~\cm\ a very narrow peak builds up for $T<T_c$ due to the
quasi-particle contribution to the conductivity.
Its intensity first increases and then diminishes as the quasi-particle number vanishes when $T\rightarrow 0$.
In Fig.~\ref{fig:lambda1}(a) the temperature dependence of $\sigma(\omega,T)$ is plotted for selected frequencies.
From room temperature down to $T=30$~K we find the THz conductivity basically not dependent on frequency, as expected for a normal metal; maybe small indications of fluctuations below 50~K can be identified. In the superconducting state, the conductivity at very low frequency ($\nu=4.6$~\cm) increases strongly right below $T_c$, it passes through a maximum $\sigma_{\rm max}$ around $14~{\rm K}\approx 0.7\,T_c$, and then drops rapidly. As the frequency increases, this peak vanishes and only a simple drop of the conductivity is observed for $T<T_c$.

The BCS theory predicts a so-called coherence peak in the electrodynamic absorption of superconductors;\cite{Tinkham96,DresselGruner02} it is considered as
a hall-mark of singlet superconductivity and most pronounced in the dirty limit and for $s$-wave symmetry of the order parameter.
In contrast to theory and observations on conventional  superconductors,\cite{Gorshunov93,Steinberg08}
in the present case the quasi-particle peak is larger in amplitude and extends to lower temperatures,
as already seen in Fig.~\ref{fig:condeps1}(a). An even larger peak has been reported\cite{Hashimoto09} from microwave experiments on K-doped BaFe$_2$AS$_2$ and extensively discussed in Ref.~\onlinecite{Schachinger09}.
It is desired to conduct further experiments on different compositions and films in order to obtain more information on  scattering mechanisms, the influence of multiple bands and cross-scattering between the bands.

According to the Ferrell-Glover-Tinkham
sum rule \cite{Tinkham96,DresselGruner02} the missing area
\begin{equation}
A = \int\left[\sigma^{(n)}(\omega)-
\sigma^{(s)}(\omega)\right]{\rm d}\omega
\end{equation}
between the  conductivity in the normal and the superconducting state is collected in the $\delta$-peak of the  condensate at $\omega=0$. It is a measure of the London penetration depth $\lambda_L=c/\sqrt{8A}=(2500\pm 700)$~\AA.
Alternatively the contribution of the superconducting carriers is probed by the permittivity $\epsilon^{\prime}$ which for low  frequencies goes as
$1-\epsilon^{\prime} \propto \left(\omega_{ps}/\omega\right)^{2}$ in excellent agreement with the experimental results plotted in Fig.~\ref{fig:condeps1}(b). Its development with temperature is displayed in the inset of Fig.~\ref{fig:lambda1}(b).
For $T\rightarrow 0$ we obtain $\lambda_L=(3600\pm 500)$~\AA\ in good agreement with the values we obtained for single crystals.\cite{Wu10,Wu10b} The spectral weight $\left(\omega_{ps}/2\pi c\right)^2 = (1.94\pm 0.1)\times10^7~{\rm cm}^{-2}$ of the superconducting condensate is in excellent agreement with the scaling relation suggested for cuprates.\cite{Uemura91,Homes04,Wu09c}
We note that taking into account the below-gap contribution to the conductivity due to quasi-particles allows us to obtain more realistic values of the superconducting density compared to reflectivity experiments.

\section{Conclusions}
The comprehensive optical investigation of \bfca) thin films
evidence the complete opening of the superconducting gap at
$2\Delta=3.7$~meV, i.e.\ $2\Delta/k_BT_c=2.1$. From our complex
transmissivity measurements we can rule out nodes in the gap. The
frequency and temperature behavior of the complex electrodynamic
response corresponds well with the predictions of the BCS theory.

\begin{acknowledgments}
Julia Braun helped with the ellipsometric measurements.
The work was supported by the RAS Program for fundamental research ``Strongly correlated electrons in solids and solid structures''.

\end{acknowledgments}

\end{document}